\documentclass[apj]{emulateapj}
\pdfoutput=1

\usepackage{subfigure}
\usepackage{textcomp}
\usepackage{amsmath}
\usepackage[colorlinks=false,pdfborder={0 0 0}]{hyperref}
\usepackage{natbib}
\usepackage{xcolor}
\usepackage{fleqn}

\begin{document}

\title{Contacts of Water Ice in Protoplanetary Disks - Laboratory Experiments}

\email{gregor.musiolik@uni-due.de}

\author{Grzegorz Musiolik}
\affil{University of Duisburg-Essen, Faculty of Physics, Lotharstr. 1-21, 47057 Duisburg, Germany}

\author{Gerhard Wurm}
\affil{University of Duisburg-Essen, Faculty of Physics, Lotharstr. 1-21, 47057 Duisburg, Germany}

\begin{abstract}
Water ice is abundant in protoplanetary disks. Its sticking properties are therefore important during phases of collisional growth. In this work, we study the sticking and rolling of 1.1 mm ice grains at different temperatures. We find a strong increase in sticking between 175 K to 200 K which levels off at higher temperatures. In terms of surface energy this is an increase with a factor of 63.4, e.g. from $\gamma = 0.0029 \rm J/m^2$ to $\gamma = \rm 0.19 J/m^2$, respectively. We also measured critical forces for inelastic rolling. The critical rolling distance is constant with a value of 0.19 mm. In view of planetesimal formation at low temperatures in protoplanetary disks, the surface energy is not larger than for silicate dust and ice aggregation will share the same shortcommings. In general, water ice has no advantage over silicates for sticking and collisional growth might not favor ice over silicates.
\end{abstract}


\section{Introduction}

Water is among the most abundant molecules in space. Depending on the ambient temperature it is gas or ice. Both have been detected in protoplanetary disks \citep{Terada2007, Ceccarelli2005, Dominik2005}. At the low pressure of protoplanetary disks, fluid water is not present.

The detection of the water snowline in protoplanetary disk is still a diffucult task. New efforts show that e.g. ALMA is capable of determining the position of the snowline in general \citep{Notsu2018}. However, the presence of ice grains was determined by IR absorption in HL Tau in the past \citep{Cohen1975}. At the same time, there is also evidence of a fast pebble growth near to the condensation lines \citep{Zhang2015}. It has often been assumed that such growth is induced by enhanced sticking properties of water ice compared to silicates. One difference in favour is the strong dipole character of water molecules.

The aggregation of solids is an early step in planet formation, where grains collide, stick together and grow. For silicates, different barriers terminate this process, e.g. the bouncing barrier \citep{Zsom2010, Kelling2014, Kruss2016, Kruss2017} or particle loss due to radial drift  \citep{Weidenschilling1977}.
Fragmentation of still larger grains still allows growth by mass transfer \citep{Teiser2009}, but might not favor rapid growth to planetesimals either. 

Does water ice improve the situation? Sublimation, sintering and condensation were considered by \citet{Saito2011}. Sublimation at the snowline might release embedded dust grains, locally enhancing the density of dust grains to grow. This principle was verified in experiments by \citet{Aumatell2011}. \citet{Ros2013} also considered the recondensation of water vapor to grow larger bodies outside of the snowline.
In the colder outer and therefore more stable regions, collisions might again be dominating
interactions. Which size scale is actually important then is not straight forward to be answered. Sublimation and recondensation might change the structure of grains and everything from nm contacts to virtually boulder size ice blocks is thinkable. 

The literature on water ice collisions is rich. \citet{Bridges1996} early on carried out slow speed (pendulum like) collisions with macroscopic bodies determining coefficients of restitution. These were often used for studies of Saturnian ring particles which are in parts composed of water ice \citep{Colwell1990}. \citet{Higa1998} and \citet{Higa1996} studied the collisonal behaviour of centimeter sized water ice spheres and found that the rings should rather consist of unsmooth ice particles. \citet{Hill2015} added data on coefficients of restitution for cm-sized spheres measured under microgravity. \citet{Deckers2015} found in high speed collisions with solid ice that mass transfer is an option to grow a larger icy body. Collision experiments with small micrometer water ice grains reveal sticking even at velocities of 10 m/s where 1 m/s is typical for silicates \citep{Gundlach2015, Poppe2000, Musiolik2016a, Musiolik2016b, Blum2008}. Finally, simulations of \citet{Dominik1997} showed that the stability of ice aggregates is increased over silicate aggregates. Therefore, a number of these works suggest that ice indeed improves the situation over silicates. 
It is therefore natural that \citet{Kataoka2013} argued that nm-ice grains might even grow to huge webs of ice beyond the water snowline of several AU in protoplanetary disks, before they get compacted, evantually. This would essentially circumvent all barriers of planetesimal formation noted before. 

However, not all these data are within the parameter range of protoplanetary disks. Water is only solid below 200 K at the low pressure of protoplanetary disks. In laboratory settings at 1 bar, temperatures essentially up to 273 K are feasible for experiments.
It might also be noted that aside from planet formation, ice in this rather high temperature range is important on Earth and many physical ice values are often restricted to or deduced from this common range of higher temperatures. 

Sticking of ice grains can be characterized by their surface energy. Typical values found are e.g. 0.19 J/m$^2$ \citep{Gundlach2011a}. For comparision, the specific surface energy between the liquid and vapor phase was measured to $0.076$ J/m$^2$ and the grain boundary (solid-solid) energy in the same work to $0.065$ J/m$^2$ \citep{Ketcham1969} for rather high ($>260 K$) temperatures. As mentioned and since water is rather volatile, it might not suffice to give just one value as surface energy might strongly depend on temperature. \citet{Gaertner2017} find that above 200 K there is an increase of the diffuse layer thickness ontop of ice grains and the surface energy might increase due to it. Recently,  \citet{Gundlach2018} suggested that the surface energy might strongly decrease below 200 K.

One of our goals here is to quantify this and to measure the static force necessary to separate an ice grain from a flat ice target and deduce the surface energy from this.
The sticking force $F_\text{JKR}$ for a spherical particle of radius $R$ with a flat target is given by the JKR contact model \citep{Johnson1971}
\begin{equation}
F_\text{JKR} = 3 \pi \gamma R
\label{fs}
\end{equation}

For multiple contacts, the expression for the total contact force changes to the sum
\begin{equation}
\begin{aligned}
F_\text{S}&=3\pi\gamma\sum_{i=1}^{N}R_i \\ 
&\approx 3\pi\gamma N R_\text{m} =3\pi\gamma N \phi R.
\end{aligned}
\label{FST}
\end{equation}

Here, $N$ is the number of total contacts and $\phi$ a correction factor which connects the curvature radius $R$ of the ideal spherical grain to a mean curvature radius of sub-contacts $R_\text{m}$.

A second goal is the determination of rolling forces necessary to initiate irrersible rolling. This quantity also enters in collisions of dust aggregates \citep{Dominik1995, Dominik1997}. \citet{Aumatell2014}, e.g., found that predictions for the relation between twisting torques and sticking did not follow the predicted behaviour for nm-size grains. So rotational motion at all relevant size scales of particles is an important process and crucial parameters are currently essentially unknown.
Irreversible rolling sets in if the tangential force $F_t$ applied in the center of the grain is \citep{Dominik1995}
\begin{equation}
F_t = 6 \pi \gamma \xi.
\label{ft}
\end{equation}
$\xi$ is the critical linear rolling distance or critical shift which is unknown. For (sub)-micrometer grains it was argued by \citet{Dominik1995} to be on the order of the atomic scales or 0.2 nm. \citet{Heim1999} determined the value to about 3 nm by atomic force microscopy for SiO$_2$ grains. This is a factor of 10 off but in view of possible values of the surface energy as compiled by \citet{Kimura2015}, this is uncertain by a factor 10. 

Combining eq. \ref{fs} and \ref{ft}, the parameter $\xi$ is independent of the surface energy and can be calculated from the measured forces

\begin{equation}
\xi = \frac{1}{2}\frac{F_t}{F_\text{JKR}} R.
\label{xieq}
\end{equation}

It is important to note that the model underlying this assumes a single contact between two spheres.
However, non-perfect spheres might form a second or third contact or even $N$ contacts as e.g.
often assumed in the context of friction between macroscopic bodies. 

\begin{figure}
	\centering
	\includegraphics[width=0.7\linewidth]{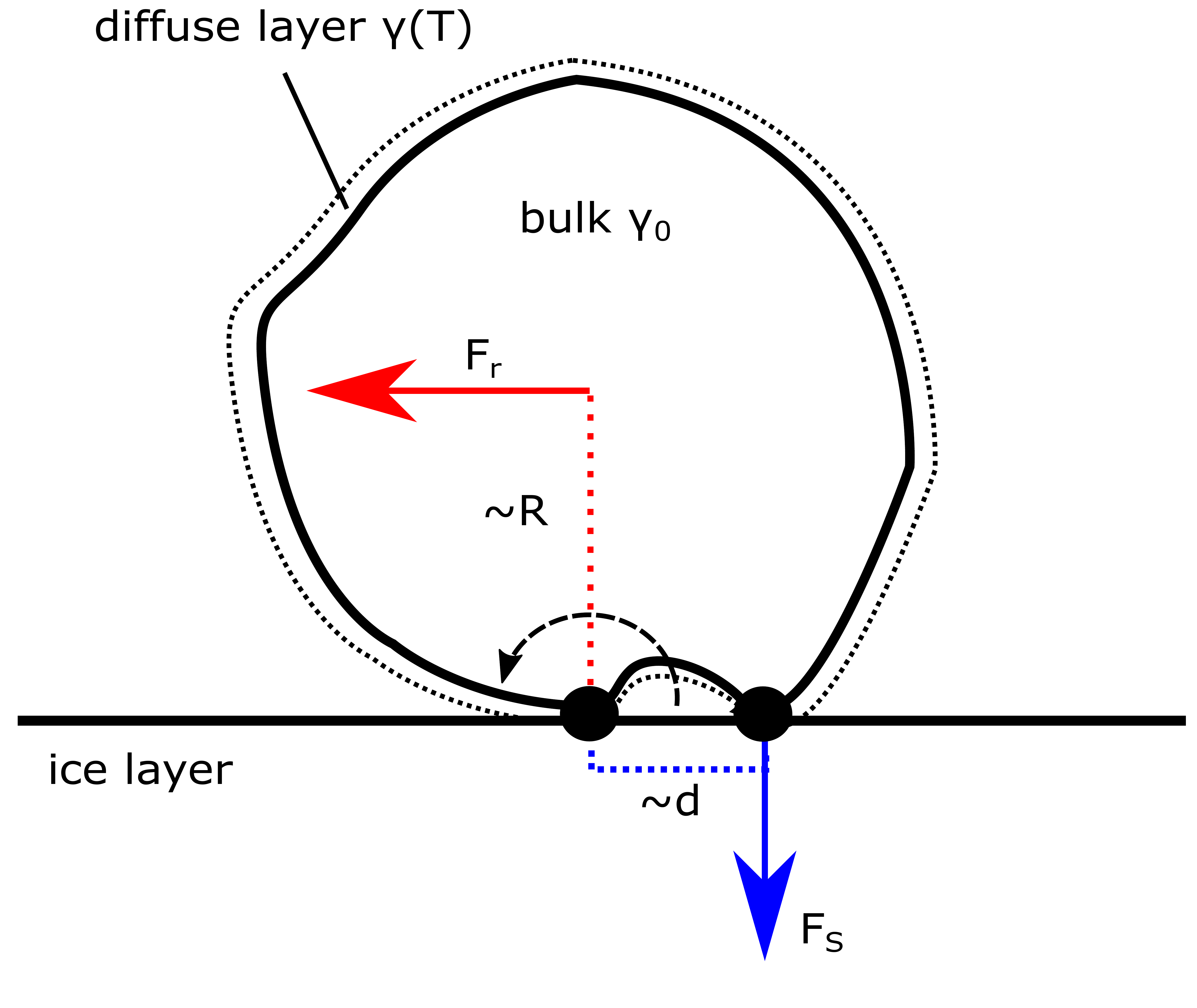}
	\caption{Rolling of a non-perfectly spherical ice grain. The grain sticks at 3 contacts (two left in line, one right) with a force of $F_\text{s}$ each in this example. The torque by the solenoid induces a rolling motion of the ice grain, which is balanced by the torque applied by the sticking of the second and third contact in distance $d$. In general, $N-N_R$ contacts will break and an mean distance $d$ has to be considered. The sphere rolls over $N_R$ sub-contacts.} 
	\label{morecontacts}
\end{figure}

If more than one contact exist some distance away from the first one, the argument for initiating rolling changes to the requirement of breaking $(N-N_R)$ of these contacts by applying the necessary pull-off force (eq. \eqref{fs}). $N_R$ is the amount of contacts over which the grains rolls. An example is sketched in fig. \ref{morecontacts}. $(N-N_R)$ of these contacts apply a mean resistive torque of $F_\text{JKR}d$.
In a similar way as eq. \eqref{xieq} we then find rolling under the condition that the applied torque equals this resistive torque or

\begin{equation}
\begin{aligned}
F_\text{r} R &= \sum_{i=N_R+1}^{N}F_\text{JKR,i} d_i + \sum_{i=1}^{N_R}F_{t,i}\xi_i \\
&\approx (N-N_R) F_\text{JKR}d. \\
&\Rightarrow d\approx \frac{1}{(N-N_R)} \frac{F_\text{r}}{F_\text{JKR}}R.
\end{aligned}
\label{d}
\end{equation}

Here, approximately a mean contact force $F_\text{JKR}$ holds for all contacts with a mean distance $d$ between the $N$ contacts. $\sum_{i=1}^{N_R}F_{t,i}\xi_i$ is negligible due to $d\gg \xi_i$.

Both - the single and the multiple contact model - result in similar equations (\ref{xieq} and \ref{d}). So independent on the model a size $\xi$ or $d$ describes rolling friction. Only the absolute values can differ by orders of magnitude between atomic scales and scales related to the particle size as detailed below. In the following, we only refer to the contact displacement $d$ as this is the only plausible model explaining our data as detailed below.

Ice might be amorphous but might also have different crystal structures \citep{Gaertner2017, Sirono2017, Ros2013}. Starting in laboratory experiments with water droplets cooled down, hexagonal ice forms which will be the basis of this work. 
In the following we describe the experiment in which we measured the sticking and rolling forces for different temperatures, deducing $\gamma$ and $d$ for mm grains.

\section{Experiments}

\begin{figure*}
	\centering
	\includegraphics[width=0.8\linewidth]{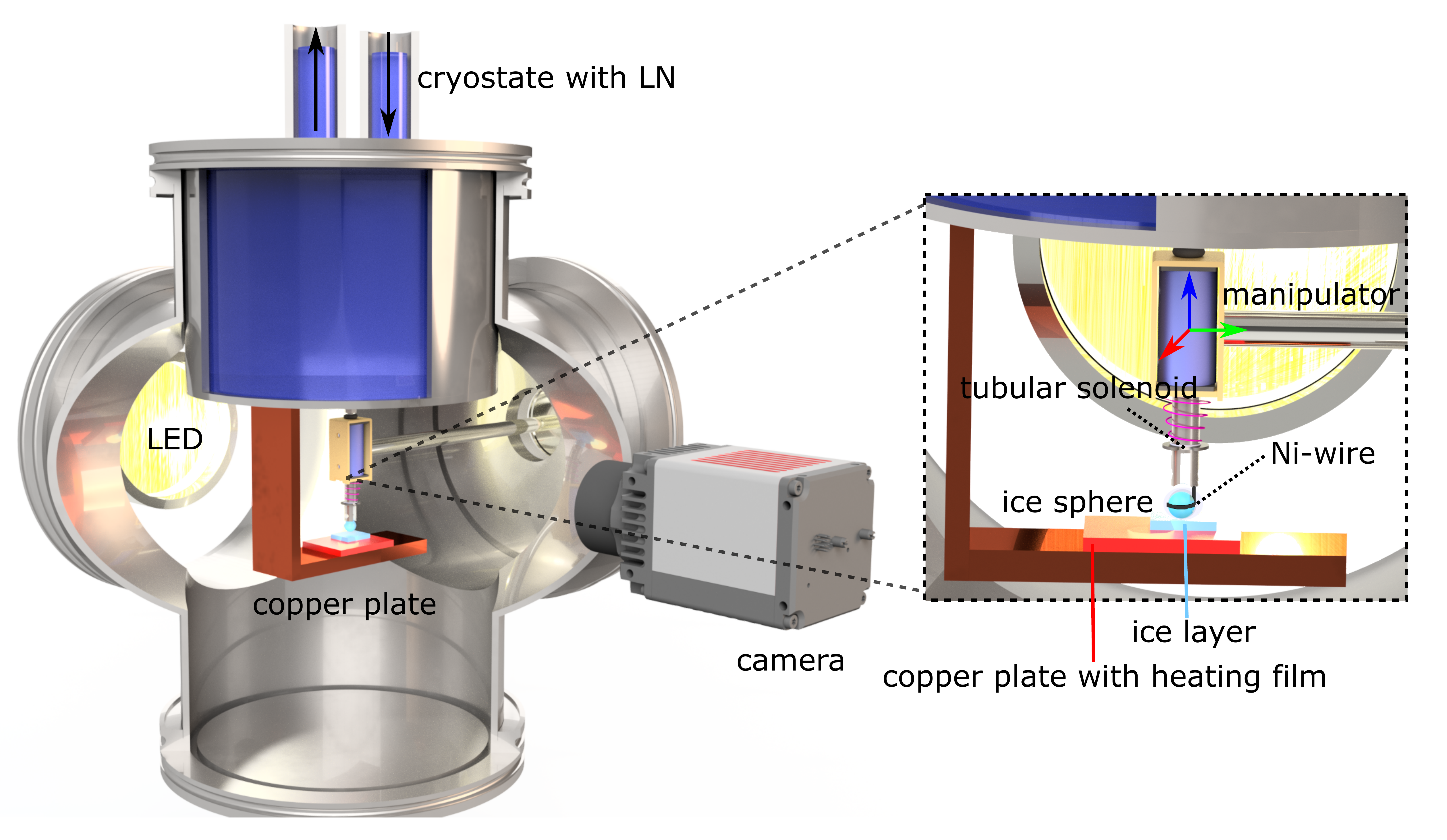}
	\caption{\label{fig.setup}Schematics of the experiment. The experiment is placed inside of a vacuum chamber at 1.5 mbar. Low temperature is achieved by cooling with liquid nitrogen inside of a cryostate. A copper plate is welded to the cryostate. With a heating film below a copper block, the temerature can be regulated in a range of 170-250 K. An ice layer is created on top of the copper block. With a manipulator, an ice sphere can be moved and rotated. Contact forces are measured optically with a high speed camera at 3000 fps by analyzing the acceleration created by a tubular solenoid. }
\end{figure*}

Fig. \ref{fig.setup} shows the setup of the experiment. A droplet is attached to a nickel wire  (diameter of 0.25 mm) with a circular loop at the one end. The droplet is placed with the help of a pipette on this wire's loop* and arranges due to the surface tension to a spherical shape. It freezes out near to a copper plate cooled by liquid nitrogen by convection and thermal radiation. 

At the other end, this wire is attached to an iron core of the mass of 2.02 g which is positioned at the inside of a tubular solenoid. Adjusting the current allows the application of small forces. The resolution limit is $0.05$ mN. The whole assembly is attached to a micromanipulator above a copper plate welded to a cryostate and can be moved and rotated in any direction. The experiment is placed inside a vacumm chamber and measurements are done at 1.5 $\pm$ 0.1 mbar. The pull off events are observed by a high speed camera at 3000 fps with bright field LED illumination.  

The temperature is regulated before each experiment with a precision of $\pm$ 1 K. A heating film counteracts the cooling of the copper plate from the cryostate on top of the vacuum chamber. The cryostate is cooled down with liquid nitrogen to a temperature af approximately 80 K. The copper plate can be cooled down by this to 170 K. Once the temperature is set, a water layer is placed on the copper plate and freezes  together with the droplet on the nickel wire. The vacuum chamber is then evacuated. The ice sphere is put in contact with the ice layer. Pull-off forces are measured for a certain temperature.

\begin{figure*}
	\centering
	\includegraphics[width=0.8\linewidth]{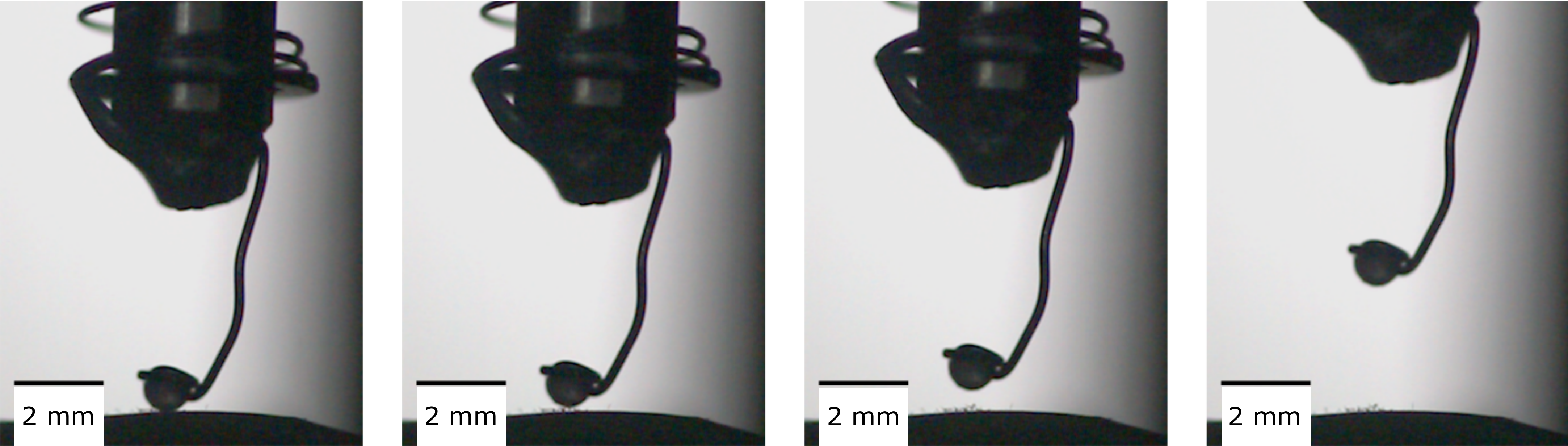}
	\caption{\label{Stickexample}Example sequence of images used for sticking force measurements. The individual images of this example are taken at 30 fps. The ice grain is accelerated by a tubular solenoid. The contact breaks at a critical force. From the acceleration this critical force can be calculated.}
\end{figure*}

Fig. \ref{Stickexample} shows an image sequence of a measurement, where the particle is pulled off its target. 
At the moment the particle contact breaks, the lifting force balances gravity and sticking force. Since the sticking force is instantaneously lost, the net force remaining is the lifting force of the same magnitude. From the acceleration of the sample holder deduced from the images and the mass of the holder known, this force can be calculated. Friction inside the coil is on the order of $0.05$ mN which is the resolution limit of any direct measurement of pull off forces. The force applied to the particle is directed upwards. Small torques may still be present but do not lead to movement (rolling) of the particle before lift in the measured data.

\begin{figure}
	\centering
	\includegraphics[width=0.6\linewidth]{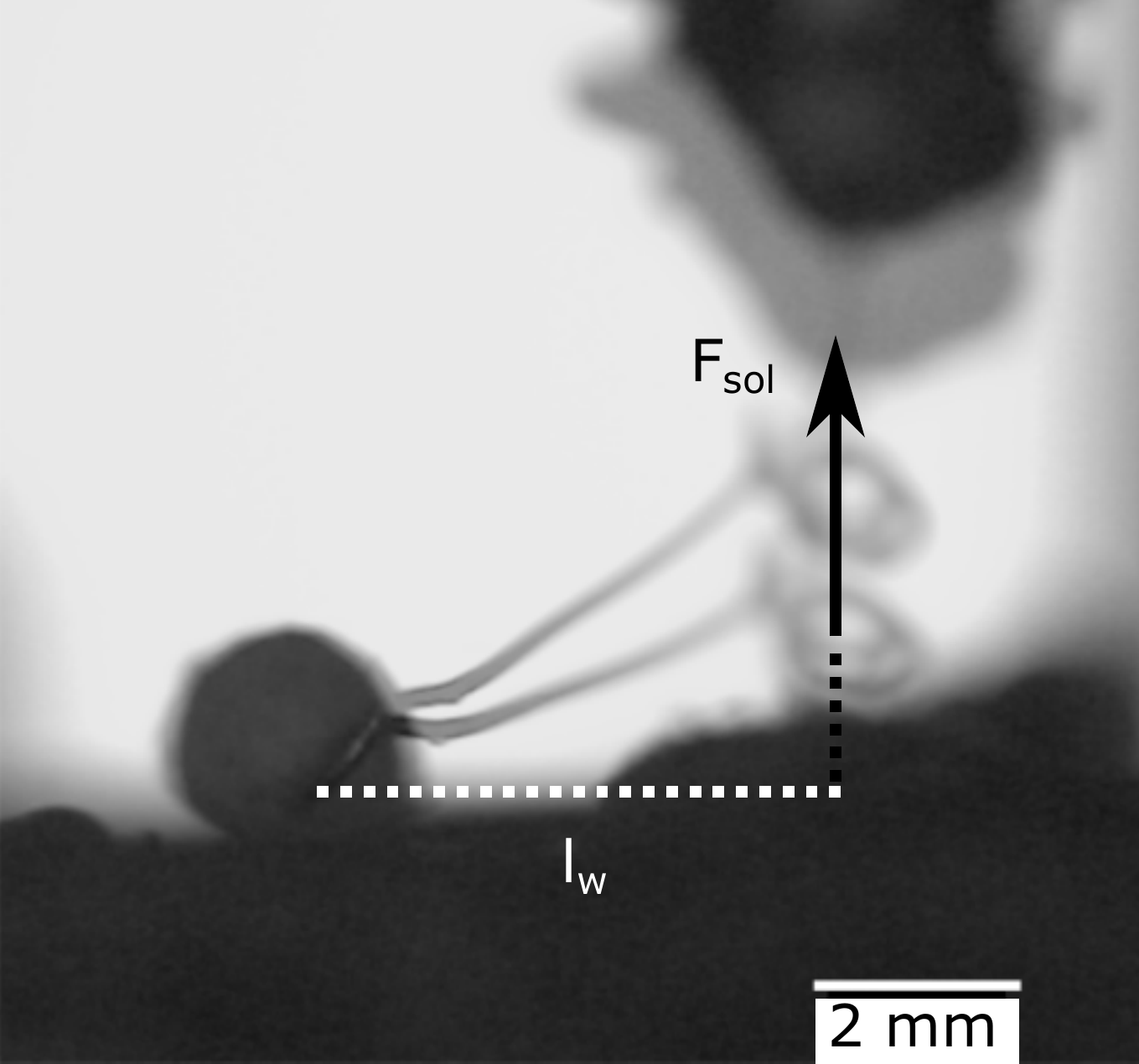}
	\caption{Example of images used for rolling force measurements. The tubular solenoid applies a force to a wire connected perpedicular to the ice sphere which results in a torque acting on the ice grain. The force on the solenoid allows to determine critical rolling forces with the proportion of the lever to the grain's radius.}
	\label{rollexample}
\end{figure}

In a similar fashion though, rolling is observed. Here, larger torques are applied on purpose by choosing a different geometry. An example is shown in fig. \ref{rollexample}.

\section{Results}

The results of these measurements are summarized in fig. \ref{gammas}. In this figure, we present the critical pull off forces, the deduced surface energies, the critical rolling forces and the contact distance.

\subsection{Sticking and Surface Energy}

\begin{figure*}
	\centerline{
	\includegraphics[width=0.49\textwidth]{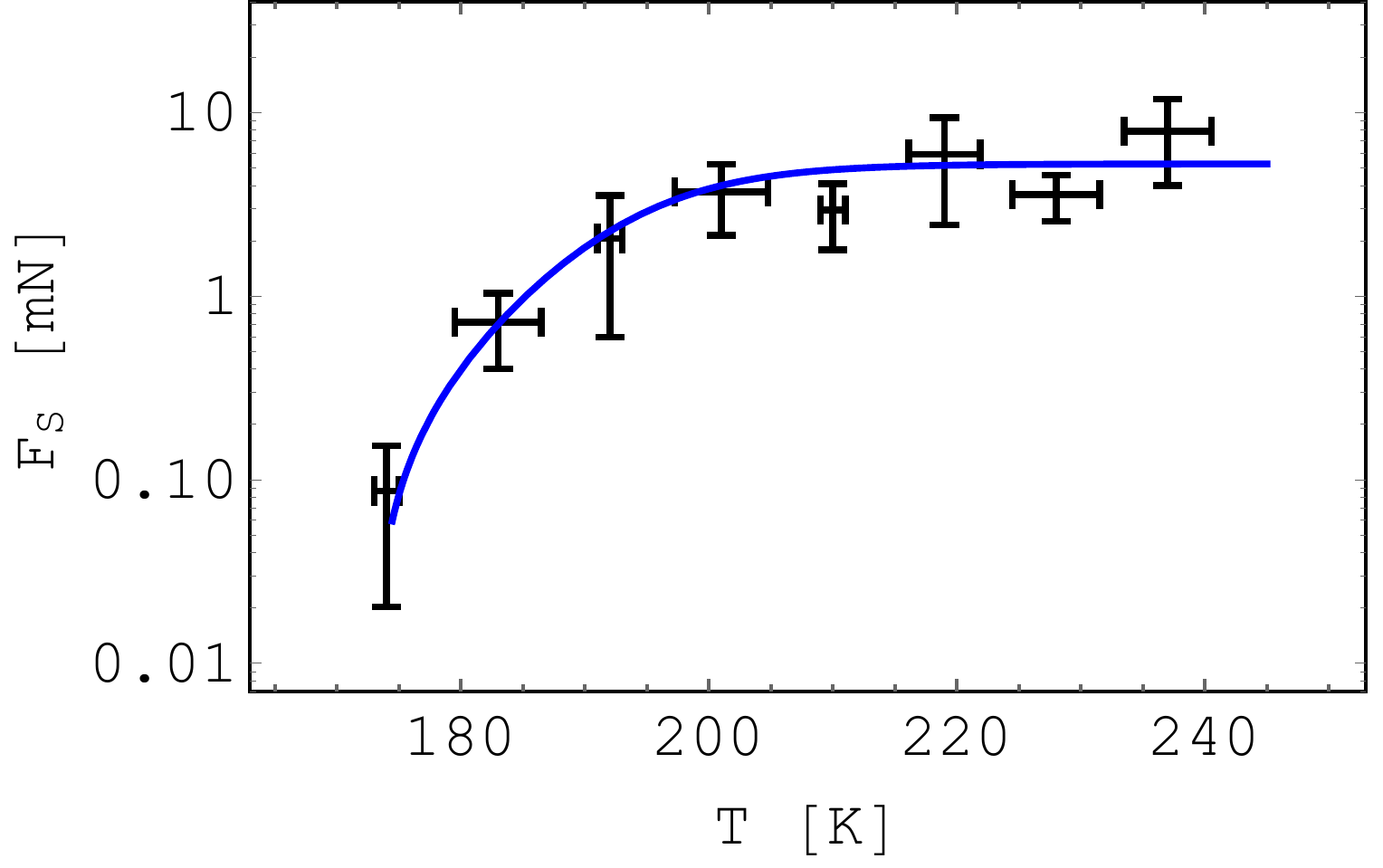}
	\includegraphics[width=0.49\textwidth]{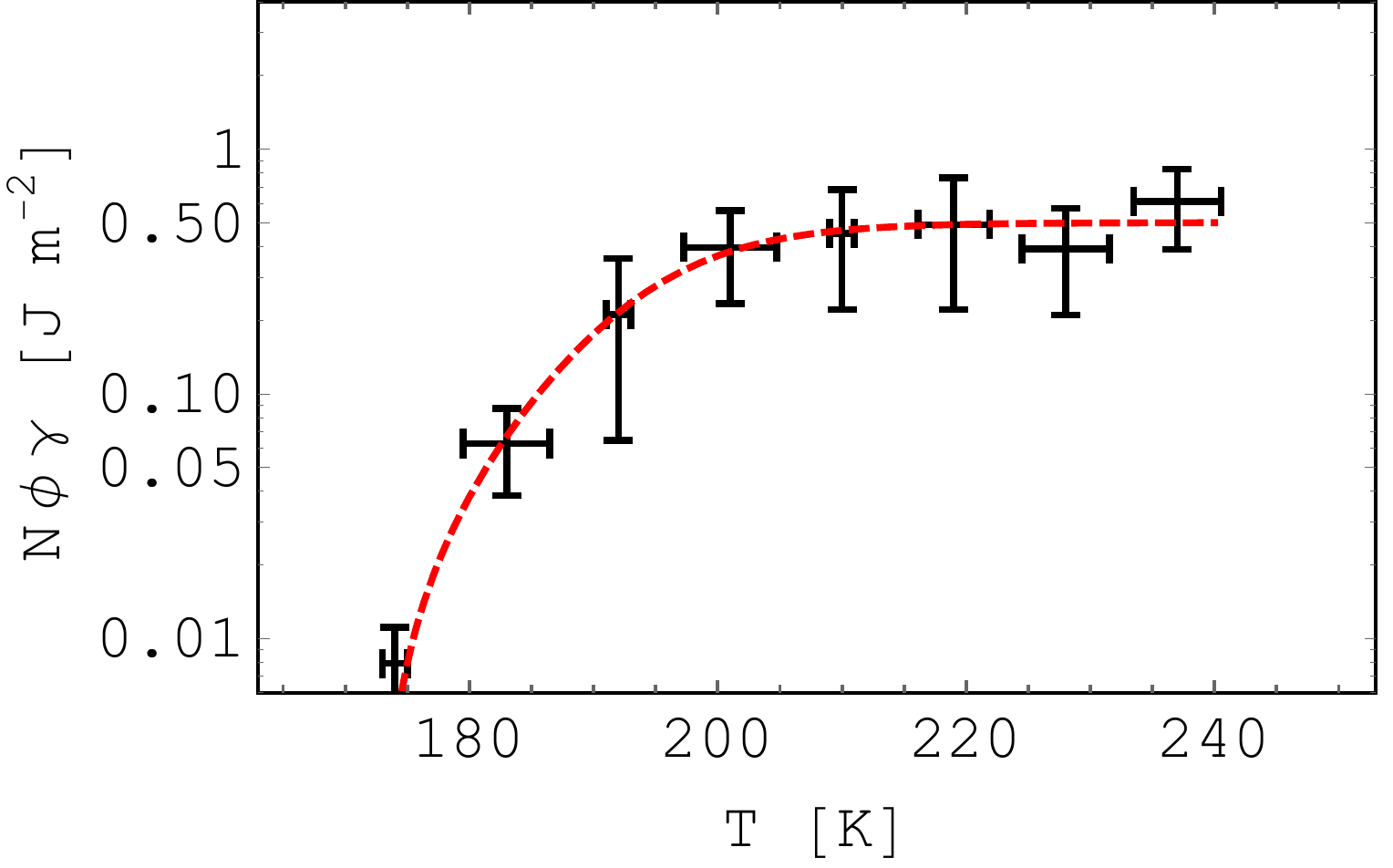}
	}
	\centerline{
	\includegraphics[width=0.49\textwidth]{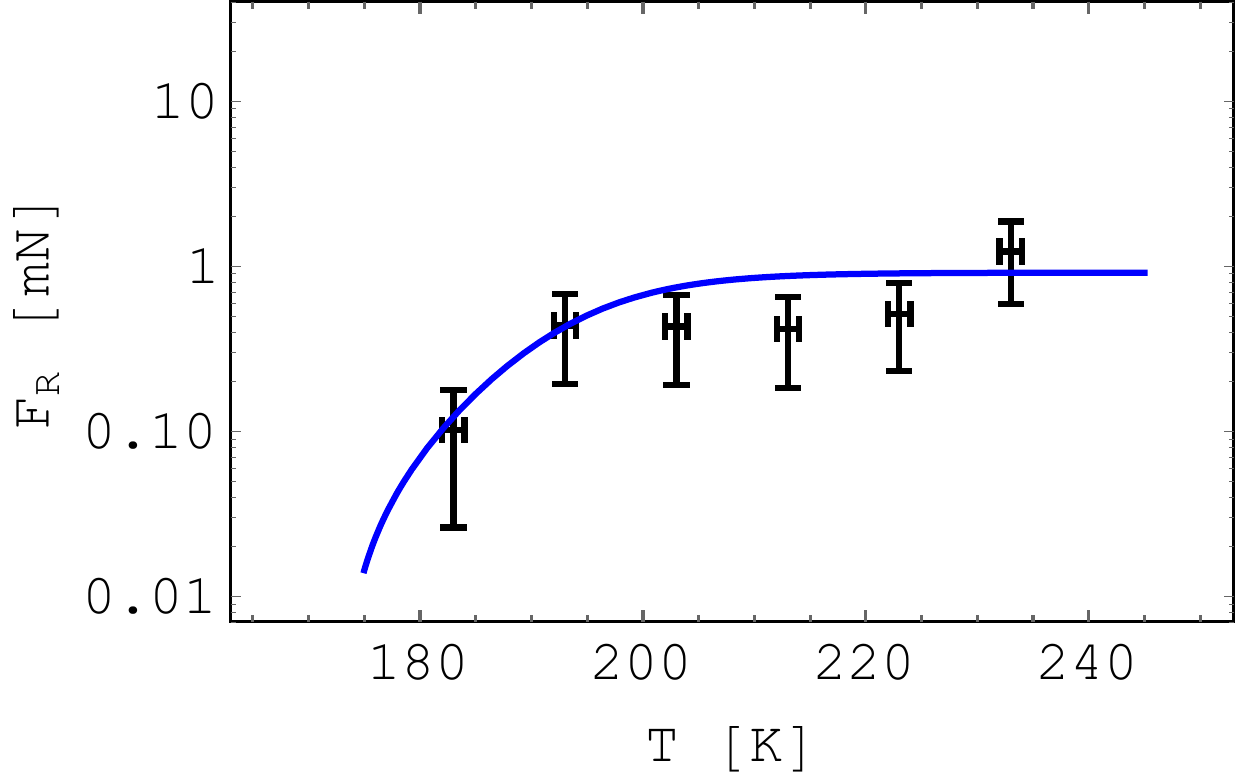}
	\includegraphics[width=0.49\textwidth]{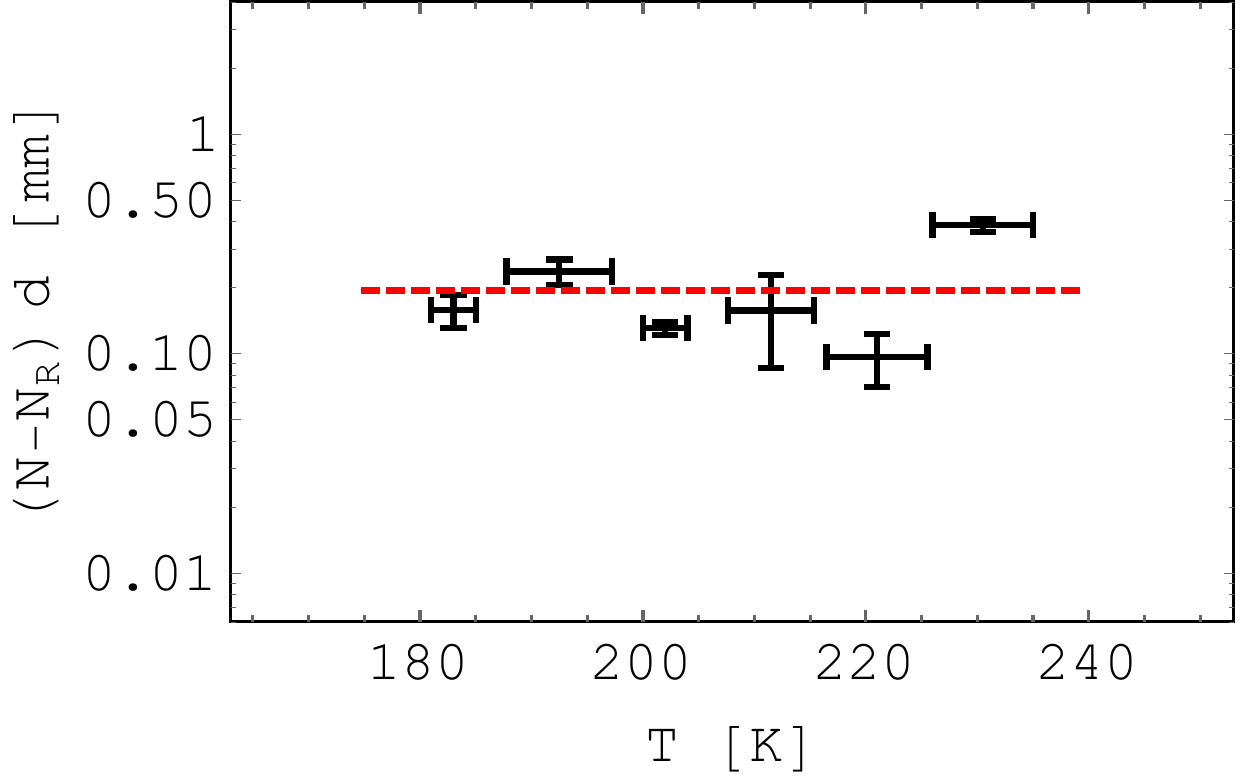}
	}
	\caption{Top, left: Sticking forces depending on temperature for a averaged grain size of 1.11 mm. The data shows 32 averaged pull-off events. The fuction for the sticking force is $3N\pi\gamma(T)\phi R$ with $R= 1.11$ mm and $\gamma(T)$ from the fit function for the surface energie of eq. \eqref{gT} (top, right of this image).
	Top, right: Surface energy deduced from the sticking forces. The value varies with the contact number $N$ and the ratio $\phi$ between the mean curvature radius of the sub contacts $R_\text{m}$ and curvature radius of the ice grain $R$. The data is fitted with the function from eq. \eqref{gT} with fitting parameters as described above.
	Bottom, left: Rolling forces depending on temperature. The plotted function is $3\pi N\phi\gamma(T)(N-N_R) d$ with the fitted $\gamma(T)$ and the mean distance $(N-N_R) d=0.194$ mm.
	Bottom, right: Critical shift depending on temperature. The values were determined by values for critical sticking and rolling forces using eq. \eqref{xieq} for the mean size $R=1.11$ mm. \newline\newline} 
	\label{gammas}
\end{figure*}

Fig. \ref{gammas} (top, left) shows the sticking force measured depending on temperature for an average grain radius of 1.11 mm. The grain sizes vary by 0.31 mm, but we essentially always measure grains of the given mm radius. The variation is not large enough to study size dependent effects in the given work. Altogether, 32 sticking events were measured and every datapoint is averaged from four values.

The sticking force increases from 0.1 mN to 5 mN between a temperature range of 175 K and 240 K.
Using eq. \ref{FST}, these forces can be transformed into surface energies using the JKR model.
These are shown in fig. \ref{gammas} (top, right).
We note that at this time, we still consider the grain to attach with $N$ contacts to the surface with an effective curvature radius of $\phi R$ for the sub-contacts.

At the high temperature beyond 200 K a constant value of 0.5 J/m$^2$ results. However, below 200 K the surface energy drops monotoneously and rapidly and reaches a value of 0.0079 J/m$^2$ at 175 K. This is more than a factor 10 lower than typically assumed. The experiment is currently limited to that temperature range so we do not know if the surface energy levels off or continues to drop to lower temperatures.

The pull-off measurements were done in less than 30 s after placing the ice grain on the ice surface. To test a possible dependence on the contact time of grain and surface we varied the contact time for a high temperature of 233 K. The time does not have an impact on the contact forces which can be assumed as constant within the first 120 s. The mean contact force is $9.93 \pm 0.57$ mN after 10 s and $7.83 \pm 1.97$ mN after 120 s for a temperature of 233 K.

We attribute the observed temperature dependence to two phenomenons. The first one is the loss of the diffuse layer on the ice droplet's surface at smaller temperatures. The diffuse layer is typically $10-20 \mathring{A}$ thick and is defined by a gradient in density \citep{Gaertner2017}. Within this diffuse layer, a mixture between amorphous ice, liquid water and H$^+$ and OH$^-$ ions can exist. The other effect is the change of the crystal structure of the ice at lower temperatures. Both effects were analyzed by \citet{Gaertner2017} in neutron scattering experiments. 
If a diffuse layer is present, the surface energy of the diffuse layer contributes strongly to the contact force. 

We assume the surface energy to be a superposition of the bulk material and $\gamma_d(T)$. In order to describe the data analytically, we choose a heuristic funtion for the surface energy
\begin{equation}
\begin{aligned}
\gamma(T) &= \gamma_c+ \gamma_{d0}\tanh\left(\beta \left(T-T_0\right)\right).
\end{aligned}
\label{gT}
\end{equation}

Eq. \eqref{gT} is fitted with the data in fig. \ref{gammas} and results in $\gamma_c = 0.0078$ J/m$^2$ and $\gamma_{d0} =0.5$ J/m$^2$, $\beta=0.078$ K$^{-1}$ and $T_0= 193.2$ K.

With eq. \eqref{gT}, also the sticking force can be described. With the mean grain size of $R=1.11$ mm, the resulting sticking force is $3N\pi\gamma(T)\phi R$. The plot is shown in fig. \ref{gammas} (top, left).

Generally, the absolute value of the surface energy can be determined at low temperatures assuming its value for high temperatures is given. This is independent of the number of contacts and effective radii as long as they are the same on average for all temperatures. For example, for a literature value for the surface energy of $\gamma=0.19$ J/m$^2$ at high temperature a surface energy of only  $\gamma = 0.0029 \rm J/m^2$ follows for $T = 175K$.
The number of contacts $N$ and the factor $\phi$ are not constrained in detail by our measurements.
However, assuming a minimum of $N=3$ contacts gives $\phi=0.83$ or $\phi R\approx0.93$ mm which would be consistent with the observation of almost spherical ice grains (s. fig. \ref{Stickexample}). 

\subsection{Rolling and Critical Shift}

The image analysis also allows to determine critical rolling forces. From the force on the solenoid $F_\text{sol}$, the force acting directly at the ice grain can be deduced to
\begin{equation}
F_\text{r}=F_\text{sol}\frac{l_\text{w}}{R}
\end{equation}
with the length of the lever $l_\text{w}$ and the grain's radius $R$. The torques on an individual contact are typically on the order of $10^{-6}$ Nm. 

For high temperatures above 200 K, the critical rolling forces are on the order of 1 mN which is 10 times lower than sticking forces. Similar to the behavior of direct sticking forces, the critical rolling forces decrease toward lower temperature. 

On first look, this ratio of 1/10 between rolling and pull-off force seems compatible to the values used by \citep{Dominik1997} for basaltic dust. 
However, the ratio there is tied to the critical shift of a single contact. Throughout the temperature range from 180 - 230 K, the critical shift would remain constant but with a large mean value of $0.19$ mm.  
This is quite different from the assumption of necessary shifts on the atomic scale as argued for micrometer size grains in \citet{Dominik1997}. This is physically unreasonable so we assume that the grain is sticking to the plate with more than one contact.  
If we consider again $N=3$ as minimum and in addition two contacts to roll over or $N_R=2$, the contact displacement would remain $d=0.19$ mm which is reasonable. 

As an extreme variation with a large number of contacts, there is a the transition from macroscopical distances $d_i$ to a microscopical roughness with a vast number of asperities. The influence of such a microscopical surface roughness can then be described by an effective surface area for the contact force, e.g. described by \citet{Greenwood1966}. In this case, the contact would again be given by the curvature radius of the grain or $N\phi=1$. We can also approximate the number $N$ by the ratio $d_0/d_i$ then or $(d_0/d_i)\phi=1$ with a constant $d_0$. Therefore, the factor $\phi$ is directly proportional to the asperity displacements and in general will be related to $d$.

The resistive rolling is a factor of about 10 lower than the sticking force and requires a value for the contact displacement $d$ of $0.19$ mm or about 10 \% of the particle's diameter. If that ratio holds for other grain sizes remains to be seen.

\section{Conclusion}

We measured the critical sticking and the rolling forces for 1.1 mm water ice spheres within the temperature range from 175 K to 240 K at a pressure of 1.5 mbar. Sticking forces increase from 0.1 mN to 5 mN within this temperature range with a saturation at 200 K which corresponds to a surface energy of 0.0029-0.19 J/m$^2$ according to the JKR contact model and assuming 3 contacts. Their distance in our case is 0.19 mm or 10\% of the particle diameter. The rolling forces are approximately a factor 10 lower than sticking forces and depend on temperature as well due to the dependence on $\gamma$.

The surface energy determines the sticking velocity, where $v_\text{stick} \propto\gamma^{(5/6)}$  \citep{Dominik1997}. As ratio between warm and cold sticking, we get $\gamma(>200K)/\gamma(T\approx175K)\approx 63.4$ and the sticking velocities decrease to lower temperatures by a factor 31.72. 

The snowline in a solar system like protoplanetary disk is at a few AU depending on the details of the system. For a special model based on the minimum mass solar nebula (MMSN) it is located approximately at 2 AU \citep{Musiolik2016a, Hayashi1981}. According to the model \citep{Hayashi1981}, the temperature at 2 AU is 198 K. Already at 2.5 AU, the averge temperature drops to 178 K. Thus, if at all, the advantage of enhanced ice sticking can only contribute to growth of larger aggregates in a very narrow location around the snowline, which at the same time is also strongly influenced by sublimation and condensation. This might explain why the growth of pebbles was observed near to condensation lines in the past \citep{Zhang2015}. 

It has been tempting to assume that water ice promotes the formation of planetesimals due to increased sticking beyond the snowline. However, at temperatures of protoplanetary disks, where water is solid, the surface energy in most cases is not larger than the surface energy for silicate dust. Therefore, the advantage of increased sticking is lost. 

\section*{acknowledgements}
This project was funded by DFG grant WU 321/12-1. 


\end{document}